\documentclass[10pt]{article}
\usepackage[OE]{express}
\usepackage{palatino}
\usepackage{tikz}
\usetikzlibrary{arrows, decorations.pathmorphing, backgrounds, positioning, fit, petri, automata}
\usepackage{color}
\usepackage{amssymb}
\usepackage{graphicx}
\usepackage{amsmath}
\usepackage{colordvi}
\usepackage{bbm}
\usepackage{cite}
\usepackage{titlesec}
\usepackage[OT1]{fontenc}
\usepackage{amsfonts}

\begin{document}

\title{Double-passage ground-state cooling induced by quantum interference
in the hybrid optomechanical system}
\author{ Lingchao Li,$^{1}$ Longjiang Liu,$^{1}$ Shuo Zhang,$^{2,\ast }$
Jian-Qi Zhang,$^{3,\# }$ Mang Feng$^{3,\dag}$}

\address{\authormark{1} College of Science, Henan University of Technology, Zhengzhou 450001, China \\
\authormark{2}Zhengzhou Information Science and Technology Institute, Zhengzhou, 450004, China\\
\authormark{3}State Key Laboratory of Magnetic Resonance and Atomic and Molecular Physics, Wuhan Institute of Physics and Mathematics, Chinese Academy of Sciences, and Wuhan National Laboratory for Optoelectronics, Wuhan, 430071, China\\ }


\email{\authormark{$\ast$}zhang31415926@gmail.com; \authormark{$\#$}changjianqi@gmail.com; \authormark{$\dag$}mangfeng@wipm.ac.cn}

\begin{abstract}
We propose a quantum interference cooling scheme for a nano-mechanical
resonator (NAMR) in a hybrid optomechanical system, where the atoms are
trapped in an optomechanical cavity, coupling to an additional
optical cavity. The absorption of the optomechanical resonator can be modified
by quantum interference effects induced by the atom-cavity and cavity-cavity
couplings independently. With the employment of the quantum interference,
the desired transition for cooling is enhanced, along with the heating suppression due to
the undesired transition. As a result, the NAMR vibration
can be cooled down to its ground state. Particularly, with the assistance of the
atoms, our scheme is experimentally feasible even for lower
qualities cavities, which much reduces the experimental difficulty.
\end{abstract}




\section{Introduction}

Cavity optomechanics works as an ideal platform to study the quantum
properties of macroscopic mechanical systems. For this reason, it has been
employed to create non-classical states \cite{pra-84-024301,prl-101-200503},
realize quantum information processing \cite{Fiore,Wang1,Dong,Schmidt,Haokun}%
, and achieve precision control and measurement \cite{LaHaye,Krause,Buchmann,Purdy}.
Since the NAMR is very sensitive to
small deformation, the precision measurement based on NAMR can approach
the Heisenberg limit \cite{Natphys-5-509}. However, this measurement is
limited by the thermal noise on NAMR. To further enhance the measurement sensibility,
it is necessary to eliminate the thermal noise and
cool the NAMR down to its ground state.

Until now, many different cooling schemes have been proposed to achieve the
ground-state cooling of NAMR\cite{Kippenberg,Marquardt}. One of the famous
cooling methods is the sideband cooling \cite{Wilson-Rae,Marquardt2,Genes1}%
, which works in the resolved-sideband regime with the decay rate of the cavity
much less than the vibrational frequency of the NAMR, as verified experimentally \cite{Riviere}.
However, the sideband cooling is hard to be used in most of the physical system, such as the NAMR,
since the decay rates are always larger than the vibrational frequency.

As such, some efforts have been made in the
nonresolved-sideband regime, which works for the decay rate larger than the
vibrational frequency of the NAMR. For example, the cooling can be realized
with a dissipative coupling \cite{Li,elste,Xuereb,Weiss1,Yan}, where the
main dissipation is employed as the coupling between the cavity and the NAMR,
and a fast ground-state cooling of the NAMR is available with time-dependent
optical driven cavities \cite{Li2}. Moreover, an essential cooling method
working in the nonresolved-sideband regime is based on quantum
interference \cite{Lin,Weis,Safavi-Naeini,Agarwal}, in which the quantum interference
is used to modify the absorption spectrum of the NAMR in the cooling of the NAMR to
the ground state. The modification of the absorption spectrum is due to destructive interference
of quantum noise \cite{elste,Xuereb}. Since there is a large decay in the cooling schemes
via quantum interference, the speed of the cooling via quantum interference can be much faster than the one of the
sideband cooling \cite{Gu}.

On the other hand, with the development of the optomechanics \cite{Aspelmeyer}, hybrid
atom-optomechanics becomes an essential branch of the optomechanics. The additional atom
modifies the quantum features of the optomechanics \cite{Pflanzer,Ian}, and realizes atom-mechanical
entanglement and quantum steering \cite{Genes,Lilingchao,Tan}. Moreover, it
has been shown theoretically and experimentally that the mechanical
ground-state cooling is available in hybrid atom-optomechanical systems
\cite{zhangshuo,Vogell,Shi}, as the absorption spectrum of the NAMR can be
modified by the interaction between the atoms and the optical field
indirectly, which allows the NAMR to be cooled down to its vibrational ground
state.

Here we propose a cooling scheme with atoms trapped in an optomechanical
cavity and coupling to a single-mode optical cavity. In our system, there
are two different channels for quantum interference, one of which is from the
atom-cavity coupling and the other of which is from the cavity-cavity coupling. Both of
them satisfy the conditions for two-photon resonance. The combination
of these two quantum interference effects can not only reduce the
experimental requirement that is limited by the cavity decay rates, but also
enhance the red sideband transition for cooling and suppress the blue
sideband one for heating. As a result, the ground-state cooling of the NAMR
can be achieved.

Compared with the previous works involving only one quantum interference effect
\cite{Guo,Liu2,Genes3}, our scheme with the additional quantum interference effect is more efficient
and can cool the NAMR down to its ground state with less mean phonon number. Moreover,
different from the previous works for cooling, which is limited by the decay
rates of the cavity \cite{Guo,Liu2} and the atoms \cite{Genes3}, our scheme can work
even with larger decay rates due to the combination of
two quantum interference effects. As a result, our scheme reduces the experimental
difficulty and works within a broader parameter regime.
In addition, the speed of the phonon dissipation can be much faster in our scheme
than in the previous works \cite{Guo,Liu2,Genes3}, due to larger decay rates and a more thorough
suppression of the transitions for heating.

The remaining part of the present paper is organized as follows. In Section
2 we describe the system model, Hamiltonian and the rate equations for
cooling. The absorption spectrum of the NAMR and cooling mechanism are given
in Section 3. The simulations and discussion are given in Section 4 and a
conclusion is presented in the last section.

\section{Model, Hamiltonian, and the rate equations for cooling}

As shown in Fig. 1, an ensemble of $N$ atoms are trapped in an
optomechanical cavity $2$ with a single atom-cavity coupling strength $g_{a}$
\cite{Bariani}. The optomechanical cavity $2$ couples to the
single-mode cavity $1$ with a strength $J$, and is also driven by an
external field at a frequency $\omega _{l}$ with a driven strength $%
\varepsilon $.\ Thus the Hamiltonian of this system can be written as ($%
\hbar =1$)\
\begin{eqnarray}
H &=&\omega _{1}a_{1}^{\dagger }a_{1}+\omega _{2}a_{2}^{\dagger
}a_{2}+\omega _{m}b^{\dagger }b+\omega _{eg}s_{ee}+i(\varepsilon
a_{2}^{\dagger }e^{-i\omega _{l}t}-\varepsilon ^{\ast }a_{2}e^{i\omega
_{l}t})  \notag \\
&&+J(a_{1}^{\dagger }a_{2}+a_{2}^{\dagger
}a_{1})+g_{a}(a_{2}s_{eg}+s_{ge}a_{2}^{\dagger })-ga_{2}^{\dagger
}a_{2}(b+b^{\dagger }),  \label{101}
\end{eqnarray}%
where $a_{j}$ $(a_{j}^{\dagger }$, $j=1,2)$ and $b$ $(b^{\dagger })$ are
annihilation (creation) operators for cavity $j$ and NAMR, which take the
corresponding frequencies $\omega _{j}$ and $\omega _{m}$, respectively;
each atom owns an excited state $\left\vert e\right\rangle $ and a
ground state $\left\vert g\right\rangle $ with a transition frequency $%
\omega _{eg}$, and $s_{eg}=\sum_{\mu =1}^{N}|e^{\mu }\rangle \langle g^{\mu
}|$ and $s_{ee}=\sum_{\mu =1}^{N}|e^{\mu }\rangle \langle e^{\mu }|$; $g$ is
the single-photon radiation coupling coefficient. The first four items in
Hamiltonian (1) are for the free Hamiltonians of the two cavities,
the NAMR and the atoms. The fifth item shows the drive on the
optomechanics. Other items describe the interactions for
cavity-cavity, atom-cavity, and radiation coupling.

We define the ground (excited) state of the atomic ensemble is $|\mathbf{{g}%
\rangle }=|g_{1},g_{2},...,g_{N}\rangle $ ($|\mathbf{{e}\rangle }=\frac{1}{%
\sqrt{N}}\Sigma _{j=1}^{N}|g_{1},g_{2},,e_{j},...,g_{N}\rangle $). So the
above Hamiltonian (\ref{101}) can be rewritten as
\begin{eqnarray}
H &=&\omega _{1}a_{1}^{\dagger }a_{1}+\omega _{2}a_{2}^{\dagger
}a_{2}+\omega _{m}b^{\dagger }b+\omega _{eg}\sigma _{ee}+i(\varepsilon
a_{2}^{\dagger }e^{-i\omega _{l}t}-\varepsilon ^{\ast }a_{2}e^{i\omega
_{l}t})  \notag \\
&&+J(a_{1}^{\dagger }a_{2}+a_{2}^{\dagger }a_{1})+\sqrt{N}g_{a}(a_{2}\sigma
_{eg}+\sigma _{ge}a_{2}^{\dagger })-ga_{2}^{\dagger }a_{2}(b+b^{\dagger }),
\label{1}
\end{eqnarray}%
with $\sigma _{ee}=|\mathbf{{e}\rangle \langle {e}|}$ and $\sigma _{eg}=|%
\mathbf{{e}\rangle \langle {g}|}$.

In the rotating frame with frequency $\omega _{l}$, the Hamiltonian (\ref{1}) changes to
\begin{eqnarray}
H &=&\Delta _{1}a_{1}^{\dagger }a_{1}+\Delta _{2}a_{2}^{\dagger
}a_{2}+\omega _{m}b^{\dagger }b+\Omega \sigma _{ee}+i(\varepsilon
a_{2}^{\dagger }-\varepsilon ^{\ast }a_{2})  \notag \\
&&+J(a_{1}^{\dagger }a_{2}+a_{2}^{\dagger }a_{1})+g_{a}(a_{2}\sigma
_{eg}+\sigma _{ge}a_{2}^{\dagger })-ga_{2}^{\dagger }a_{2}(b+b^{\dagger }),
\label{2}
\end{eqnarray}
with detunings $\Delta _{1,2}=\omega _{1,2}-\omega _{l}$ and $\Omega =\omega
_{eg}-\omega _{l}$.
\begin{figure}[tbp]
\centering
\includegraphics[bb=44 185 599 420,scale=0.55]{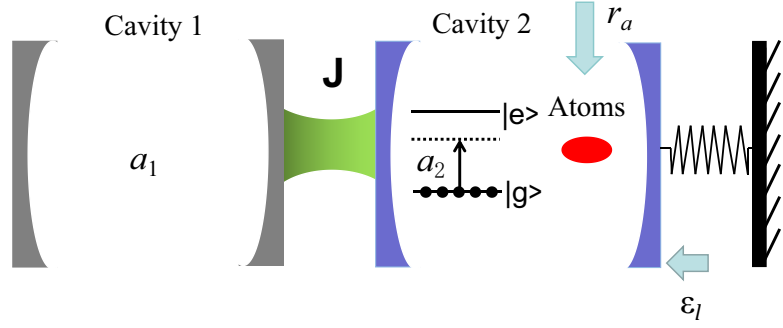}
\caption{Schematic of our scheme. The optical cavity 1 couples to the
optomechancial cavity 2 with the strength $J$. Some two-level atoms, initially
populated in ground states, are trapped in the optomechancial cavity with a coupling strength $%
g_{a}$. The optomechancial cavity is also driven by an external field $%
\protect\varepsilon _{l}$. The right-hand side mirror of the optomechancial cavity can
be moved through radiation pressure. }
\end{figure}

The dynamics of this system is governed by following quantum Langevin equations,
\begin{eqnarray}
\dot{a}_{1} &=&-(\kappa _{1}+i\Delta _{1})a_{1}-iJa_{2}+\sqrt{2\kappa _{1}}%
a_{1,in},  \notag \\
\dot{a}_{2} &=&-(\kappa _{2}+i\Delta _{2})a_{2}-iJa_{1}-i\sqrt{N}g_{a}\sigma
_{ge}+iga_{2}(b+b^{\dagger })+\varepsilon +\sqrt{2\kappa _{2}}a_{2,in},
\notag \\
\dot{b} &=&-(\gamma _{m}+i\omega _{m})b+iga_{2}^{\dagger }a_{2}+\sqrt{%
2\gamma _{m}}b_{in},  \notag \\
\dot{\sigma}_{ge} &=&-(\gamma +i\Omega )\sigma _{ge}-i\sqrt{N}g_{a}a_{2}+%
\sqrt{2\gamma }f_{in},  \label{3}
\end{eqnarray}%
where $\kappa _{j},\gamma _{m}$ and $\gamma $ denote the decay rates of the
cavity $j$, NAMR and atomic ensemble, respectively. $O_{in}$ is the noise
operator with a nonzero correlation function $\left\langle
O_{in}(t)O_{in}^{\dagger }(t^{\prime })\right\rangle =\delta (t-t^{\prime })$%
, for $O=a_{1,2}$, and $f$. $b_{in}$ is the quantum Brownian noise operator
of the NAMR with correlation functions $\left\langle b_{in}^{\dagger
}(t)b_{in}(t^{\prime })\right\rangle =n_{m}\delta (t-t^{\prime })$, and $%
\left\langle b_{in}(t)b_{in}^{\dagger }(t^{\prime })\right\rangle
=(n_{m}+1)\delta (t-t^{\prime })$, where the thermal phonon number $%
n_{m}=[\exp (\omega _{m}/(k_{B}T))-1]^{-1}$ is dependent on an environment
temperature $T$ with $k_{B}$ being the Bolzmann constant.

Suppose that the atomic ensemble is initially in its the ground state $\left\vert
\mathbf{g}\right\rangle $ \cite{zhouling}. Thus the steady-state values of
the system are%
\begin{eqnarray}
&&\left\langle a_{1}\right\rangle =-\frac{iJ\left\langle a_{2}\right\rangle
}{\kappa _{1}+i\Delta _{1}},~\left\langle a_{2}\right\rangle =\frac{%
\varepsilon }{\kappa _{2}+i\tilde{\Delta}_{2}+\frac{g_{a}^{2}N}{\gamma
+i\Omega }+\frac{J^{2}}{\kappa _{1}+i\Delta _{1}}},  \notag \\
&&\left\langle b\right\rangle =\frac{ig\left\vert \left\langle
a_{2}\right\rangle \right\vert ^{2}}{\gamma _{m}+i\omega _{m}},~\left\langle
\sigma _{ge}\right\rangle =\frac{ig_{a}N\left\langle a_{2}\right\rangle }{%
\gamma +i\Omega },  \notag
\end{eqnarray}%
with $\tilde{\Delta}_{2}=\Delta _{2}-g(\left\langle b\right\rangle
+\left\langle b^{\dagger }\right\rangle )$. In the case of $g_{a}=0$, our
system will be reduced to the model in Ref.\cite{Guo}.

With the application of the linearization approximation \cite%
{Wilson-Rae,Marquardt}, all the operators can be written as the sum of
steady-state mean values and their fluctuations, e.g., $a_{1,2}=\left\langle
a_{1,2}\right\rangle +\delta a_{1,2}$, $b=\left\langle b\right\rangle
+\delta b$, $\sigma _{ge}=\left\langle \sigma _{ge}\right\rangle +\delta
\sigma _{ge}$. Then the above Langevin equations are rewritten as%
\begin{eqnarray}
\delta \dot{a}_{1} &=&-(\kappa _{1}+i\Delta _{1})\delta a_{1}-iJ\delta a_{2}+%
\sqrt{2\kappa _{1}}a_{1,in},  \notag \\
\delta \dot{a}_{2} &=&-(\kappa _{2}+i\tilde{\Delta}_{2})\delta
a_{2}-iJ\delta a_{1}-i\sqrt{N}g_{a}\delta \sigma _{ge}+iG(\delta b+\delta
b^{\dagger })+\sqrt{2\kappa _{2}}a_{2,in},  \notag \\
\delta \dot{b} &=&-(\gamma _{m}+i\omega _{m})\delta b+iG(\delta
a_{2}^{\dagger }+\delta a_{2})+\sqrt{2\gamma _{m}}b_{in},  \notag \\
\delta \dot{\sigma}_{ge} &=&-(\gamma +i\Omega )\delta \sigma _{ge}-i\sqrt{N}%
g_{a}\delta a_{2}+\sqrt{2\gamma }f_{in},
\end{eqnarray}%
with $G=g\left\langle a_{2}\right\rangle $.

The corresponding effective Hamiltonian for the above Langevin equations is given by
\begin{equation}
H_{eff}=H_{ceff}+\omega _{m}\delta b^{\dagger }\delta b-g(\delta
a_{2}^{\dagger }+\delta a_{2})(\delta b+\delta b^{\dagger })  \notag
\end{equation}%
with the effective Hamiltonian for cooling%
\begin{equation}
\begin{array}{ccl}
H_{ceff} & = & \Delta _{1}\delta a_{1}^{\dagger }\delta a_{1}+\tilde{\Delta}%
_{2}\delta a_{2}^{\dagger }\delta a_{2}+\Omega \delta \sigma _{ee} \\
& + & J(\delta a_{1}^{\dagger }\delta a_{2}+\delta a_{2}^{\dagger }\delta
a_{1})+\sqrt{N}g_{a}(\delta a_{2}\delta \sigma _{eg}+\delta \sigma
_{ge}\delta a_{2}^{\dagger }).%
\end{array}
\label{5}
\end{equation}%
The Langevin equations of Hamitonian (\ref{5}) for cooling are presented as%
\begin{equation}
\begin{array}{ccl}
\delta \dot{a}_{1} & = & -(\kappa _{1}+i\Delta _{1})\delta a_{1}-iJ\delta
a_{2}+\sqrt{2\kappa _{1}}a_{1,in}, \\
\delta \dot{a}_{2} & = & -(\kappa _{2}+i\tilde{\Delta}_{2})\delta
a_{2}-iJ\delta a_{1}-i\sqrt{N}g_{a}\delta \sigma _{ge}+\sqrt{2\kappa _{2}}%
a_{2,in}, \\
\delta \dot{\sigma}_{ge} & = & -(\gamma +i\Omega )\delta \sigma _{ge}-i\sqrt{%
N}g_{a}\delta a_{2}+\sqrt{2\gamma }f_{in}.
\end{array}
\label{6}
\end{equation}
In the weak coupling regime, the back action of the NAMR can be ignored.
With the radiation force defined as $F=\delta a_{2}^{\dagger }+\delta a_{2}$, the
absorption spectrum for this force is $S_{FF}(\omega )=\int
dte^{i\omega t}\left\langle F(t)F(0)\right\rangle $. So the absorption
spectrum $S_{FF}(\omega )$ of the radiation force can be calculated from the
quantum Langevin equations (\ref{6}) as%
\begin{equation}
S_{FF}(\omega )=\frac{1}{|H(\omega )|^{2}}\left( 2\kappa _{2}+\frac{%
2J^{2}\kappa _{1}}{|\kappa _{1}-i(\omega -\Delta _{1})|^{2}}+\frac{%
2Ng_{a}^{2}\gamma }{|\gamma -i(\omega -\Omega )|^{2}}\right) ,  \label{z8}
\end{equation}%
with%
\begin{equation}
H(\omega )=\kappa _{2}-i(\omega -\tilde{\Delta}_{2})+\frac{J^{2}}{\kappa
_{1}-i(\omega -\Delta _{1})}+\frac{Ng_{a}^{2}}{\gamma -i(\omega -\Omega )}.
\label{9}
\end{equation}%
The three items in Eq. (\ref{z8}) correspond to the correlation functions
from the cavity $2$, cavity $1$, and atoms, respectively.

Following the method in Refs.\cite{Wilson-Rae,Clerk}, the rate equation for
the phonon on the NAMR is given by
\begin{equation}
\begin{array}{ccc}
\dot{P}_{n} & = & (G^{2}S_{FF}(\omega _{m})+\gamma
_{m}(n_{m}+1))(n+1)P_{n+1}+(G^{2}S_{FF}(-\omega _{m})+\gamma
_{m}n_{m})nP_{n-1} \\
& - & [G^{2}S_{FF}(\omega _{m})n+G^{2}S_{FF}(-\omega _{m})(n+1)+\gamma
_{m}(n_{m}+1)n+\gamma _{m}n_{m}(n+1)]P_{n},%
\end{array}
\label{04}
\end{equation}%
where $P_{n}$ is the probability for the NAMR in the Fock state $\left\vert
n\right\rangle $.

With the application of the rate equation (\ref{04}), the final mean phonon
number of the NAMR is
\begin{equation}
\begin{array}{ccl}
n_{f} & = & \frac{\gamma _{m}n_{m}+\gamma _{c}n_{c}}{\gamma _{m}+\gamma _{c}}
\\
& \simeq  & \frac{\gamma _{c}n_{c}}{\gamma _{m}+\gamma _{c}}(\gamma
_{m}n_{m}\ll \gamma _{c}n_{c})%
\end{array}%
,
\end{equation}%
where
\begin{equation*}
\gamma _{c}=G^{2}[S_{FF}(\omega _{m})-S_{FF}(-\omega _{m})],
\end{equation*}%
is the cooling rate,
\begin{equation}
n_{c}=\frac{S_{FF}(-\omega _{m})}{S_{FF}(\omega _{m})-S_{FF}(-\omega _{m})},
\notag
\end{equation}%
is the final mean phonon number under the ideal condition.

\begin{figure}[tbp]
\centering
\includegraphics[bb=24 401 468 753,scale=0.6]{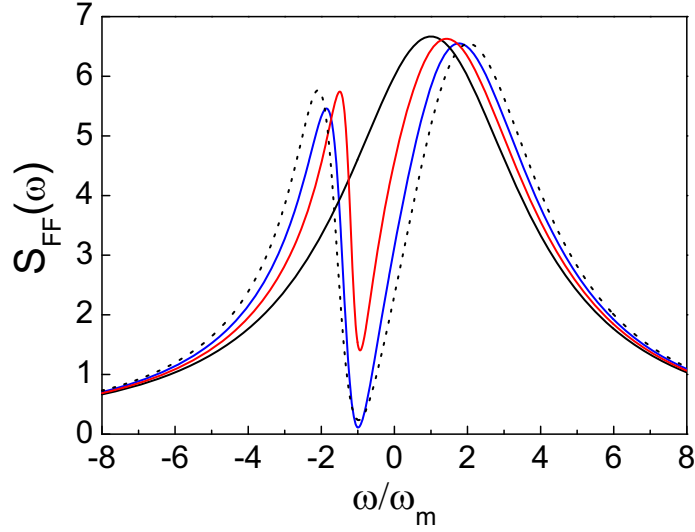}
\caption{Optical fluctuation spectrum $S_{FF}(\protect\omega )$ versus
the frequency $\protect\omega $ for the coupling coefficients $(J$, $g_{a})$, with
black solid line ($J=0$, $g_{a}=0$), red solid line ($J=\protect\omega _{m}$, $g_{a}=0$),
blue solid line ($J=0$, $g_{a}=0.1\protect\omega _{m}$), and black dotted line 
($J=\protect\omega _{m}$, $g_{a}=0.1\protect\omega_{m}$). The effective detuning
of the first cavity mode $\Delta _{1}=\Omega =-\protect\omega _{m}$ with the
decay rate $\protect\kappa _{1}=0.1\protect\omega _{m}$, while the second
cavity mode $\tilde{\Delta} _{2}=\protect\omega _{m}$ with its decay rate $%
\protect\kappa _{2}=3\protect\omega _{m}$. The number of atomic ensemble and
atomic decay rate is $N=200$, $\protect\gamma =0.1\protect\omega _{m}$,
respectively.}
\label{2}
\end{figure}

\section{absorption spectrums and cooling processes}

From now on we discuss the absorption spectrum and the cooling processes in
our system.

In Fig. 2, we show the absorption spectrum $S_{FF}(\omega )$ versus the
frequency $\omega $ with different coupling strengths ($J$, $g_{a}$). Since
the optomechanical cavity works in the nonresolved-sideband regime, i.e.,
the decay rate of cavity is much larger than the frequency of the NAMR, when
the optomechanical cavity $2$ is decoupled from the cavity 1 and the atom
ensemble ($J=0$, $g_{a}=0$), the absorption spectrum is in a Lorentz profile
with a half width $\kappa _{2}$. On the other hand, with the assistance of
the good cavity and the atomic ensemble, the Lorentz profile can be modified to a
Fano one for quantum interference \cite{Li3}, and the ground-state
cooling of the NAMR can be achieved with employment of quantum interference.

Quantum interference can be followed from the eigen-energies for
Hamiltonian (\ref{6})
\begin{equation}
\begin{array}{ccl}
E_{\pm } & = & \frac{1}{2}(\Omega +\tilde{\Delta}_{2}\pm \sqrt{%
4(J^{2}+g_{a}^{2}N)+(\Omega -\tilde{\Delta}_{2})^{2}}) \\
E_{0} & = & \Omega .%
\end{array}
\label{11}
\end{equation}%
These three eigen-energies correspond to three inflection points in
the absorption spectrum $S_{FF}(\omega)$ [see Fig. 2]. $E_{0}=\Omega$ is
for a dark state, meaning no absorption at this point for the
quantum destructive interference. As a result, we can set
\begin{equation}
\Delta _{1}=-\omega _{m}\text{ and }\Omega =-\omega _{m}  \label{10}
\end{equation}
to suppress the transition for heating process. Moreover, to get the
ground-state cooling of the NAMR, the value $S_{FF}(\omega =\omega _{m})$ in
the absorption spectrum for cooling must be enhanced. In other words, it is
necessary to ensure $E_{-}=\omega_{m}$, so that the absorption spectrum $%
S_{FF}(\omega =\omega _{m})$ can reach its maximum value by adjusting the
coupling strengths ($J$, $g_{a}$) with quantum constructive interference.
Then, we get the optimal condition for two coupling strengths as%
\begin{equation}
J^{2}+g_{a}^{2}N=2\omega _{m}(\omega _{m}-\tilde{\Delta}_{2})\geqslant 0,
\label{13}
\end{equation}%
which is independent of the decay rates, since the eigen-energy has no
relation with the decay rates.
\begin{figure}[tbp]
\centering
\includegraphics[bb=41 431 491 770,scale=0.6]{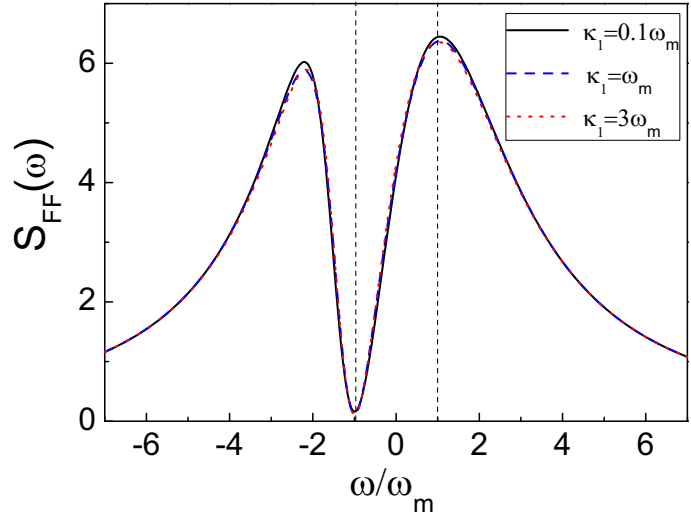}
\caption{Optical fluctuation spectrum $S_{FF}(\protect\omega )$ versus $%
\protect\omega $ for different decay rates, i.e., $\protect\kappa _{1}=0.1%
\protect\omega _{m}$ (solid line), $\protect\kappa _{1}=\protect\omega _{m}$
(dashed line) and $\protect\kappa _{1}=3\protect\omega _{m}$ (dotted line).
Other parameters are given by $g_{a}=0.1\protect\omega _{m}$, $N=200$, $%
\protect\gamma =0.1\protect\omega _{m}$, $\protect\kappa _{2}=3\protect%
\omega _{m}$, $\tilde{\Delta}_{2}=-0.1\protect\omega _{m}$, $\Delta _{1}=-%
\protect\omega _{m}$.}
\end{figure}

Figure 3 shows the absorption spectrum $S_{FF}(\omega )$ versus $\omega $
for the decay rate $\kappa_{1}$ of the cavity $a_{1}$ with parameters
of $\tilde{\Delta}_{2}=-0.1\omega _{m}$ and $J=0.45\omega _{m}$. Under the
optimal condition as Eq. (\ref{13}), the heating process can be completely
suppressed [i.e., the value of $S_{FF}(\omega =-\omega _{m})$ approaches
zero], the absorption spectrum value $S_{FF}(\omega =\omega _{m})$ for the
cooling process can reach its maximal value of the curve with a good cavity (%
$\kappa _{1}=0.1\omega _{m}$). With the increase of the cavity decay rate $%
\kappa _{1}$, a ground-state cooling can still be achieved with a bad cavity
($\kappa _{1}=3\omega _{m}$), since the absorption spectrum for the bad
cavity is almost same as the one for the good cavity. It results from the fact that
the heating process is suppressed by the quantum interference for the
atom-cavity coupling. As such, different from the methods in Refs. \cite%
{Guo,Liu2,Genes3}, where the good cavity \cite{Guo,Liu2} (atom \cite{Genes3})\
is an essential condition to guarantee the minimal value $S_{FF}(-\omega _{m})$
approaching zero, our scheme, combining two quantum interference effects, works without this condition.

With the assistance of quantum interference, the cooling mechanism of
the hybrid optomechanical system in our scheme can be understood from Fig.
4, where $\left\vert n_{1}\right\rangle$, $\left\vert n_{2}\right\rangle $
and $\left\vert n_{b}\right\rangle$ are the states of the
single-mode cavity $1$, optomechanical cavity $2$ and the NAMR,
respectively. If the system is initially in the state $\left\vert \mathbf{%
g},n_{1},n_{2},n_{b}\right\rangle$, under the action of the optical pump
field, the photons are injected into the optomechanical cavity$\ 2$, the state $%
\left\vert \mathbf{g},n_{1},n_{2},n_{b}\right\rangle $ will evolve to the
state $\left\vert \mathbf{g},n_{1},n_{2}+1,n_{b}\right\rangle $. In this
situation, the photons in optomechanical cavity $2$ are transferred into
the NAMR via the transition $\left\vert \mathbf{g},n_{1},n_{2}+1,n_{b}\right%
\rangle \rightarrow \left\vert \mathbf{g},n_{1},n_{2},n_{b}+1\right\rangle $
by the radiation coupling $G$. This heating transition can be cancelled by
two quantum destructive interference effects. One is caused by the
cavity-cavity coupling interaction $J$ via the transition from the state $%
\left\vert \mathbf{g},n_{1},n_{2}+1,n_{b}\right\rangle $ to the state $%
\left\vert \mathbf{g},n_{1}+1,n_{2},n_{b}\right\rangle $, and then to $%
\left\vert \mathbf{g},n_{1},n_{2}+1,n_{b}\right\rangle $ with an additional
phase $\pi $; the other is induced by the atom-cavity coupling $\sqrt{N}
g_{a}$, which is from the state $\left\vert \mathbf{g},n_{1},n_{2}+1,n_{b}
\right\rangle $ to the state $\left\vert \mathbf{e},n_{1},n_{2},n_{b}\right%
\rangle $, and then to the state $\left\vert \mathbf{g},n_{1},n_{2}+1,n_{b}%
\right\rangle $ with a phase $\pi$. These two quantum destructive
interference effects can suppress the transition from $\left\vert \mathbf{g}%
,n_{1},n_{2}+1,n_{b}\right\rangle \ $to $\left\vert \mathbf{g}%
,n_{1},n_{2},n_{b}+1\right\rangle $. The cooling processes are explained in
the following flowchart and Fig. 4.

\begin{tikzpicture}[->,>=stealth',shorten >=1pt,auto,node distance=2.8cm,
                    semithick]
  \tikzstyle{every state}=[rectangle, minimum width = 2cm, minimum height=1cm,text centered, draw = black, fill = white!40]

  \node[state]         (S1) at (-7.5, 0)              {$\left\vert \mathbf{g},n_{1},n_{2},n_{b}+1\right\rangle$};
  \node[state]         (S2) at (-4, 0)              {$\left\vert \mathbf{g},n_{1},n_{2}+1,n_{b}\right\rangle$};
  \node[state]         (xin1) at (-0.5, 1.25)            {$\left\vert \mathbf{e},n_{1},n_{2},n_{b}\right\rangle$};
  \node[state]         (xin4) at (-0.5, -1.25)           {$\left\vert \mathbf{g},n_{1}+1,n_{2},n_{b}\right\rangle$};
  \node[state]         (DC) at (3, 0)           {$\left\vert \mathbf{g},n_{1},n_{2},n_{b}\right\rangle$};

  \path  (S1)
               edge[left=26]              node [above] {G} (S2)
	    (S2) edge[left=26]              node  [above] {$g_a$} (xin1)
               edge[left=26]             node  [above] {$\kappa_2$} (DC)
    		    edge[right=26]             node  [above] {$J$} (xin4)
        (xin1) edge  node {$\gamma$} (DC)
        (xin4) edge  node {$\kappa_1$} (DC);
\end{tikzpicture}

\begin{figure*}[tbp]
\centering
\includegraphics[bb=2 200 724 544,clip=true,width=10.9cm]{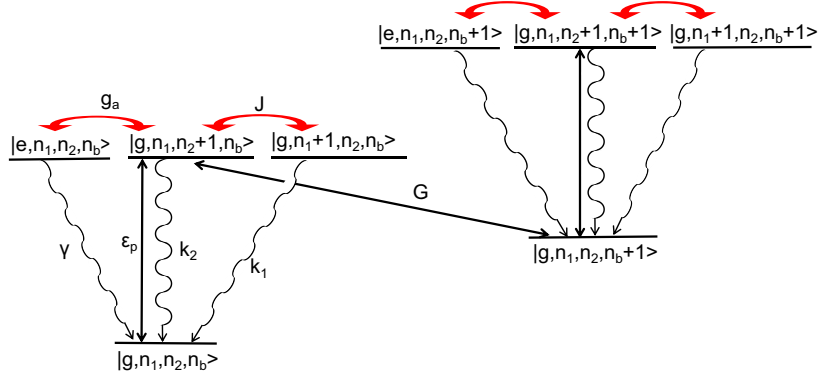}
\caption{ Level scheme for the cooling
mechanism. Here $\left\vert n_{1}\right\rangle $, $\left\vert
n_{2}\right\rangle $ and $\left\vert n_{b}\right\rangle $ denote the states
of the optical and optomechanical cavity and the oscillator, respectively. $%
\left\vert g\right\rangle $ and $\left\vert e\right\rangle $ denote the
atomic ground and excited states, respectively.}
\end{figure*}

\section{Simulation and discussion}
Next, we will verify the cooling effect of our scheme with numerical
simulations. These parameters for simulations are the follows \cite%
{Anetsberger}: $\omega _{m}=2\pi \times 20$ MHz, $Q_{m}=\omega _{m}/\gamma
_{m}=8\times 10^{4}$, $g=1.2\times 10^{-4}\omega _{m}$, $\left\vert
\varepsilon \right\vert =6000\omega _{m}$, $T=300$ mK), $\Delta _{1}=\Omega
=-\omega _{m}$ and $\kappa _{2}=3\omega _{m}$.

The final mean phonon numbers $n_{c}$ ($n_{f}$) versus the cavity-cavity
coupling strength $J$ with good cavity and bad cavity are simulated in Fig.5
(a) and (b), respectively. From the black line in Fig. 5(a), we find that the final mean phonon number $n_{c}$ can approach
zero with the increase of the cavity-cavity coupling strength $J$ [the
lowest phonon number is $n_{c}=0.18$] when the atomic ensemble is decoupled from the optomechanical cavity ($g_{a}=0$) as in Refs.\cite{Guo,Liu2}. 
It's due to the fact that, with the increase of the cavity-cavity coupling strength $J$, the line width of the spectrum
for non-absorption is enhanced, and the 2nd-order transition for
heating can be suppressed largely in this way. While there would be a dip in
the curves for the final phonon number $n_{f}$ versus the cavity-cavity
coupling strength $J$ after the environmental heating is included [the lowest
phonon number is $n_{c}=0.32$]. It can be understood as follows. With the
increase of the coupling strength $J$, the eigen-energy $E_{-}$ will deviate
from $E_{-}=\omega _{m}$, and thereby the cooling rate will decrease.
However, this problem can be overcome by introducing the atom-cavity
coupling [see Fig. 5(b)]. In this situation, the additional quantum
interference caused by the atoms can further modify the absorption spectrum, and 
the final mean phonon number $n_{c}$ turns to be very close to $%
n_{c}=0$ [see black line in Fig. 5(b)]. It means that the hybrid
optomechanical system in our scheme can cool the NAMR to its ground state
more effectively due to combination of the two quantum interference
effects, and thus the experimental difficulties are reduced in this way.

\begin{figure}[tbp]
\centering
\includegraphics[bb=47 466 545 765,scale=0.6]{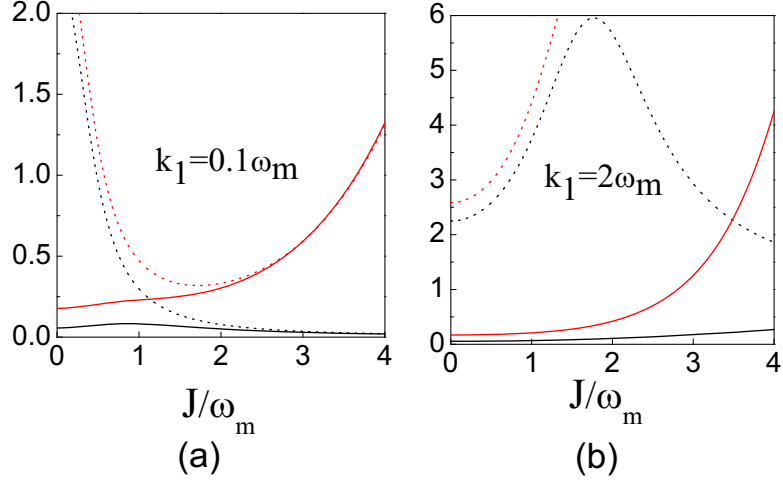}
\caption{Quantum cooling limit $n_{c}$ (black line) and
the final phonon number $n_{f}$ (red line) versus the
coupling coefficient $J$ for different decay rates of the auxiliary cavity (a) $%
\protect\kappa _{1}=0.1\protect\omega _{m}$ and (b) $\protect\kappa _{1}=2%
\protect\omega _{m}$. The dotted lines denote the pure optomechanical
system, i.e., $N=0$, and the solid lines denote the hybrid optomechanical
system, i.e., $N=100$. Other parameters are chosen as $\protect\kappa %
_{2}=3\protect\omega _{m}$, $\protect\gamma =0.1\protect\omega _{m}$, $%
g_{a}=0.1\protect\omega _{m}$, $\tilde{\Delta}_{2}=\Omega =-\protect\omega %
_{m}$.}
\end{figure}

\section{Conclusion}

In summary, we have shown the possibility of a ground-state cooling of the NAMR in the hybrid
optomechanical system, where the two-level atoms are trapped in the
optomechanical cavity, coupling to an additional optical cavity. Due to
combination of two quantum interference effects from the atom-cavity
coupling and cavity-cavity coupling, the heating processes are suppressed, while the cooling processes are enhanced.
Besides, compared with previous cooling methods involving only one quantum interference effect \cite%
{Guo,Liu2,Genes3}, the combination of two quantum interference effects can
reduce the limit on the line-widths of the cavity and atoms. As a result, our scheme can cool
the NAMR down to its ground state more efficiently. In particular, our scheme is
experimentally feasible for lower-quality cavities, which much reduces the experimental difficulty.

\section*{\textbf{Acknowledgement}}

This work is supported by the National Natural Science Foundation of China
(NSFC) (Grant Nos. 91636220 and 11674360, 11547113), the Talent Introduction Fund (No.
2016BS019) and the Youth Support Plan of Nature Science Fundamental Research
(No. 2017QNJH21) at Henan University of Technology.

\end{document}